\documentclass{appolb}
\usepackage{epsfig}

\begin{document}
\title{Total cross-section at LHC from minijets and soft gluon
resummation in the infrared region%
\thanks{Contribution to the Proceedings of 
EURIDICE Final Meeting, August 24-27th, 2006,  Kazimierz, Poland}%
}
\author{G. Pancheri
\address{INFN 
Frascati National Laboratories, P.O. Box 13, I00044 Frascati, Italy}
\and
R. Godbole
\address{Centre for High Energy Physics, Indian Institute of Science\\
Bangalore, 560 012, India}
\and A. Grau
\address{Departamento de F\'\i sica Te\'orica y del Cosmos, Universidad de 
Granada, Spain}
\and Y.N. Srivastava
\address{Physics Department and INFN, University of Perugia,
             Perugia, Italy.}
}
\maketitle
\begin{abstract}
A model for total cross-sections 
incorporating QCD jet cross-sections and soft gluon resummation is 
described and compared with present data on 
$pp$ and ${\bar p}p$ cross-sections. 
Predictions for LHC are presented for different parameter sets. 
It is shown that they differ 
according to the small x-behaviour of available parton 
density functions.
\end{abstract}

\begin{flushright}
{hep-ph/0703174}\\
{IISc/CHEP/4/07}\\
{LNF - 07 / 6(P)}\\
\end{flushright}

\section{Introduction}
The upcoming measurements at LHC are renewing considerable interest 
regarding predictions for total 
cross-sections. The model \cite{corsetti,ggps}
 we shall describe in the following, 
attempts to 
link the rate with which total cross-sections rise, to   
the infrared behaviour of the strong coupling constant 
$\alpha_s$  and to QCD hard parton-parton scattering, using 
 known phenomenological entities such as the available QCD parton density 
functions (PDFs). 

\section{The model}
The energy behaviour of the total cross-section exhibits the following 
properties \cite{PDG}
\begin{itemize}
\item an initial decrease
\item a sharp change in curvature occurring somewhere between  
20 and 50 GeV in the c.m. of the scattering hadrons
\item a smooth rise which asymptotically follows a   
$\ln s$ or $\ln^2 {s}$ type 
increase in consonance with the Froissart bound \cite{froissart,blockfroissart}
\end{itemize}
The model we use is based on
\begin{enumerate}
\item hard component of scattering responsible for the rise of the total 
cross-section \cite{allhalzen,minijets}
\item soft gluon emission from scattering particles which softens 
the rise \cite{corsetti}
\item eikonal transformation which implies 
multiple scattering and requires 
impact parameter distributions inside scattering particles and basic 
scattering cross-sections \cite{durand}
\end{enumerate}

According to our model, soft gluon emission is responsible for  the initial 
decrease in $p p $, as well as for the transformation of the sharp rise due
to the increase in gluon-gluon interactions
into 
a smooth behavior.
Thus soft gluon emission plays a crucial role, with care taken to 
extend resummation to the zero energy modes, in complete analogy for what 
is  required by the Bloch-Nordsieck 
theorem for QED\cite{BN}. 
The model can then be referred to as the BN model, for reasons which will 
also be clearer in the following.

\subsection{Details of the BN model} 
We use the following eikonal expression for the total inelastic 
cross-section:
\begin{equation}
\label{siginel}
\sigma_{inel}=\int d^2 {\vec b} [1-e^{-n(b,s)}]
\end{equation}
where $n(b,s)$ corresponds to the average number of inelastic 
collisions at  any given value of the impact parameter b. Neglecting the
real part of the eikonal, we then calculate the total cross-section as
\begin{equation}
\label{sigtot}
\sigma_{total}=2 \int d^2 {\vec b} [1-e^{-n(b,s)/2}]
\end{equation}
In our model 
$n(b,s)$
is split as
\begin{equation}
n(b,s)=n_{soft}(b,s)+n_{hard}(b,s)
\end{equation}
where we postulate the following factorization 
\begin{equation}
n_{soft/hard}(b,s)=A_{BN}^{soft/hard}(b,s) \sigma_{soft/hard}(s)
\end{equation}
with 
\begin{equation}
A_{BN}(b,s)=N \int d^2 K_{\perp}\  e^{-iK_\perp\cdot b}
 {{d^2P(K_\perp)}\over{d^2 K_\perp}} 
\end{equation}
where N is a normalization factor such that $\int d^2 {\vec b} A(b) =1$  and
\begin{equation}
\label{dpdk}
 {{d^2P(K_\perp)}\over{d^2K_\perp}} = {{1}\over{(2\pi)^2}}\int d^2 {\vec b}\ 
e^{iK_\perp\cdot b -\int_0^{q_{max}} d^3{\bar n}(k)[1-e^{-ik_t\cdot b}]}
\end{equation}
is the transverse momentum distribution 
of initial state soft gluons emitted in the  
parton-parton collisions
and where, for simplicity, 
$k_\perp\cdot b = {\vec k}_\perp\cdot {\vec b}$. In Equation (\ref{dpdk})
$q_{max}$ is the maximum transverse momentum allowed by kinematics to
single soft gluon emission in a given hard collision, 
averaged over the parton 
densities. 
According to the basic
ansatz of the Eikonal Minijet Model (EMM), 
\begin{equation}
\label{sigmajet}
\sigma_{hard}\equiv \sigma^{AB}_{\rm jet} (s) = 
\int_{p_{tmin}}^{\sqrt{s/2}} d p_t \int_{4
p_t^2/s}^1 d x_1 \int_{4 p_t^2/(x_1 s)}^1 d x_2 \sum_{i,j,k,l}
f_{i|A}(x_1) f_{j|B}(x_2)~~
 \frac { d \hat{\sigma}_{ij}^{ kl}(\hat{s})} {d p_t}.
\end{equation}
 Here   $A$ and $B$ denote particles ($\gamma, \ p,
\dots$), $i, \ j, \ k, \ l$ are parton types and $x_1,x_2$ the fractions
of the parent particle momentum carried by the parton. $\hat{s} =
x_1 x_2 s$  and $\hat{ \sigma}$ are hard parton scattering
cross--sections. As discussed in \cite{corsetti}, kinematical considerations
suggest \cite{greco}
\begin{equation}
\label{qmaxav}
q_{max}(s)={{\sqrt{s}} 
\over{2}}{{ \sum_{i,j}\int {{dx_1}\over{ x_1}}
f_{i|A}(x_1)\int {{dx_2}\over{x_2}}f_{j|B}(x_2)\sqrt{x_1x_2} \int_{z_{min}}^1
 dz (1 - z)}
\over{\sum_{i,j}\int {dx_1\over x_1}
f_{i|A}(x_1)\int {{dx_2}\over{x_2}}f_{j|B}(x_2) \int_{z_{min}}^1 (dz)}}
\end{equation}
with $z_{min}=4p_{tmin}^2/(sx_1x_2)$ and $f_{i/a}$ the valence 
quark densities used in the jet cross-section calculation.   
The steps we  follow to compare the model with data  are then the following:
\begin{enumerate}
\item choose the parameters for the hard scattering part, namely
\begin{description}
\item{(i)} parton densities (PDF), 
$p_{tmin}$ and $\Lambda_{QCD}$ for the
chosen PDF set in Equation (\ref{sigmajet}) 
\item{(ii)} model for $\alpha_s$ in the infrared region and relevant 
parameters in Equation (\ref{dpdk})
\end{description}
\item calculate $q_{max}(s,p_{tmin})$ for the given densities and
  $p_{tmin}$
using Equation (\ref{qmaxav})
\item calculate $n_{hard}(b,s)= A_{BN}^{hard}(b,s)
  \sigma_{jet}(s,p_{tmin})$
 \item choose the parameters for the low energy part, namely
\begin{description}
\item{(i)}  the constant low energy cross-section $\sigma_0$
\item {(ii)}  values for  $q_{max}^{soft}$
\end{description}
\item  calculate $n_{soft}(b,s)=
 A_{BN}^{soft}(b,s) \sigma_0(1+\epsilon {{2}\over{\sqrt{s}}})$ with
 $\epsilon=0, 1$ depending upon the process being $p p$ or $p 
\bar{p}$
\item calculate $n(b,s)$ and thus $\sigma_{tot}$ 
\item choose the parameter set which gives the best description of the
  total cross-section up to the Tevatron data \cite{e710,cdf,e811}
\end{enumerate}
Notice that once a good set of parameters is found, one can 
use $n(b,s)$ with fitted parameters to calculate survival 
probabilities or diffractive Higgs production.
\subsection{Application to  total cross-section data}
We show in this section the application of the model to  the
total 
cross-section for different PDFs. We find that our model is 
flexible enough to be able to reproduce the present data for 
$\sigma_{tot}$ using all presently available PDFs. 
In particular, all GRV \cite{GRV,GRV94,GRV98} and MRST \cite{MRST}
 densities give a good description  using 
the singular $\alpha_s$ model described in \cite{corsetti}, 
while CTEQ densities 
\cite{CTEQ} give an acceptable 
fit  up to Tevatron data, but  then fail to rise further. 

We present these results by following the previously listed steps. We start
by choosing the parameters for the hard scattering, and, following our
previous results \cite{ggps}, we fix $p_{tmin}=1.15 \ GeV$ in the jet
cross-sections and calculate   $q_{max}$  for different PDF sets.
 We show the result in Fig.(1).
\begin{figure}
\label{qmax}
\includegraphics[width=\textwidth,height=\textwidth
]{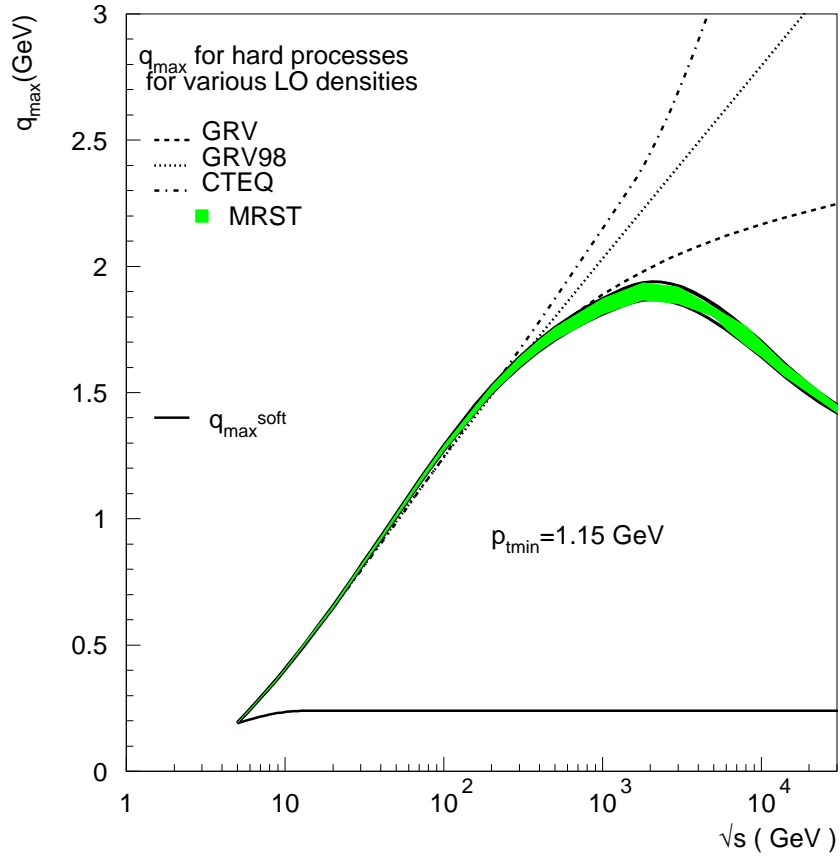}
\caption{Average value of the maximum transverse momentum allowed to single
  gluon emission, according to the model in {\protect \cite{ggps}}.}
\end{figure}
We notice the following:\begin{itemize}
\item GRV densities are of two types, GRV98\cite{GRV98}  
for which $q_{max}$ keeps on increasing 
logarithmically and the older ones 
\cite{GRV} for  which 
$q_{max}$ slows down past the TeV region, albeit still slowly increasing
\item CTEQ densities give values for $q_{max}$ which  increase more
  rapidly than $\ln {s}$ after typical Tevatron energies 
\item MRST densities indicate a behaviour opposite to CTEQ, since they give
$q_{max}$ values decreasing after the TeV cross mark.
\end{itemize} 
We now turn to the jet cross-sections and examine
the  growth with energy of 
 $\sigma_{jet}$   for different PDFs. 
 In Fig.(2)
we plot
these cross-sections for the same set of densities used to calculate
$q_{max}$
and for  $p_{tmin}=1.15 \ GeV$.
\begin{figure}
\label{sigjets}
\includegraphics[width=\textwidth,height=\textwidth
]{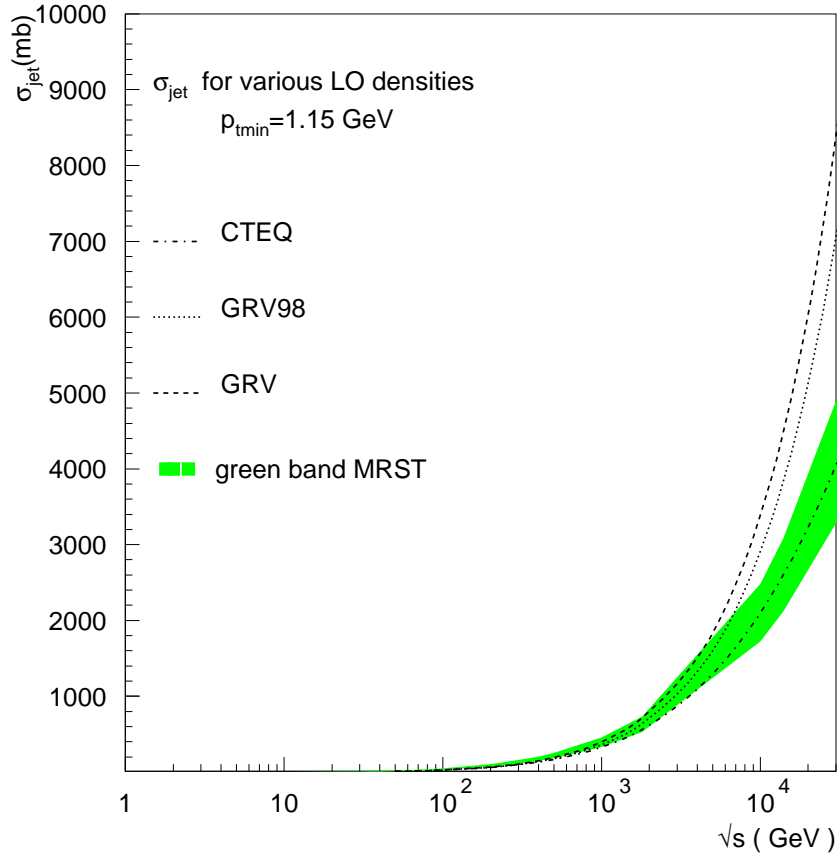}
\caption{Mini-jet QCD cross-sections for different PDFs as indicated in the
  figure}
\end{figure}
From this figure we notice that :
\begin{itemize}
\item the jet cross-sections  for GRV densities
  increase faster than all the others
\item the jet cross-sections for CTEQ increase more or less similarly to
  those for the MRST group
\end{itemize}
The implications are that $\sigma_{jet}$ with GRV densities, 
which increase faster than $\sigma_{jet}$ with MRST,
need more softening, $\sigma_{jet}$ with CTEQ, which increase 
less than with GRV, should not be
smeared that much.

To proceed further, we now need to calculate the $b-distribution$ and 
fix the low-energy parameters, like $\sigma_0$. The $b-distribution$ requires
to input the behaviour of $\alpha_s$ in the infrared region. We have shown
in \cite{ggps} the need to use a singular but integrable expression for 
$\alpha_s$ in order to reproduce both the sudden rise and the subsequent
softening of the total cross-section. Our choice is an expression like
\begin{equation}
\label{alphaRich}
\alpha_s(k_t)={{12 \pi }\over{(33-2N_f)}}{{p}\over{\ln[1+p({{k_t}
\over{\Lambda}})^{2p}]}}
\end{equation}
which depends on the singularity parameter $p$, in addition to the scale
$\Lambda$. In \cite{ggps} we have chosen the value $p=0.75$ and 
$\Lambda=100\ MeV$, other choices
are also possible \cite{pramana}. Turning now to the low-energy part, $n_{soft}(b,s)$, we
choose $\sigma_0=48\ mb$ and use for $A_{BN}^{soft}$ a set of 
$q_{max}$ values which
reproduce the low energy behaviour, and which appear in Fig. (1).  
\section{Comparison with data and expectations at LHC densities}
\begin{figure}[t]
\label{sigtotsing}
\includegraphics[width=\textwidth,height=\textwidth]{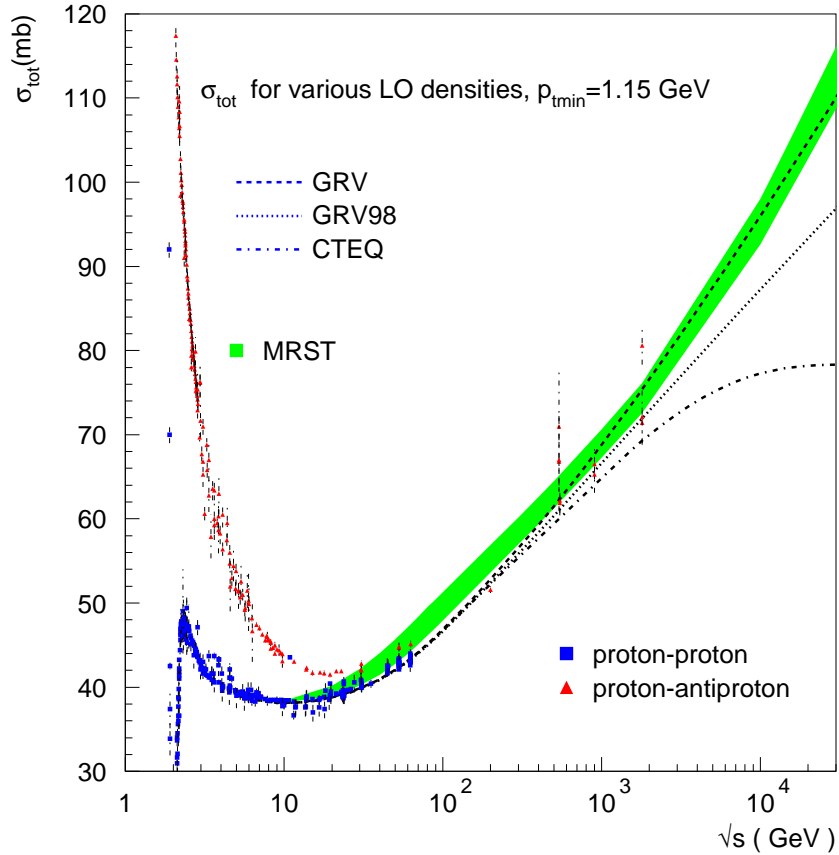}
\caption{Data for total cross-sections for $pp$ and $p{\bar p}$ 
compared with model predictions at LHC for $p p$ scattering,
 using different PDFs.}
\end{figure}

We can now input all the above in the eikonal representation for the total
cross-section and obtain
the results shown in Fig.(\ref{sigtotsing})
 for the singular $\alpha_s$ case
 and GRV, MRST and CTEQ 
densities. For the sake of clarity, we only plot curves for 
$p p$ scattering, referring the reader to 
\cite{ggps,pramana} for the curves for $p{\bar p}$ and for a different
 parameter set,  or for predictions from 
other models \cite{blockfroissart,dl,bghp,cudell,luna}. 
From the figure we see that the calculation with CTEQ densities 
appears very unlikely. The effect is due to the fact that the jet
cross-sections in the CTEQ case do not rise as much as the others 
while the softening
effect is stronger, as it is driven by $q_{max}$, which is strongly 
increasing for these densities. As a
 result, the cross-section starts decreasing. 
Notice that while the behaviour of $\sigma_{jet}$ is dominated by the 
gluon densities, that of $q_{max}$ is determined only by the valence 
quarks, as we assume this to be the leading order effect. 

The curves shown in Fig.(\ref{sigtotsing}) indicate that the coming
measurement at $\sqrt{s}=900\ GeV$ will be very important in determining
which of these curves  best describes the data. It   can then be used to 
select  the parameter set, basically PDF's and $p_{tmin}$, 
for a prediction at the  project LHC energy, $\sqrt{s}=14\ TeV$.   
If the UA5 \cite{ua5} value  at $\sqrt{s}=900\ GeV$ 
is confirmed with a comparable error, 
then, for the set of parameters discussed in this note,  at
$\sqrt{s}=14\ TeV$ our model gives
$\sigma_{total}^{GRV98}=90.2\  mb$, $\sigma_{total}^{GRV}=100.2\  mb$
and $\sigma_{total}^{MRST76}=103.4\  mb$. As shown in \cite{pramana}, 
changing the parameter set, namely $\sigma_0$, $p_{tmin}$ or the singularity 
index $p$, give values in the range $88\div 111\ mb$.

\section{Conclusions}
We have presented a version of the eikonal minijet model which allows a
good description of total cross-sections at asymptotic energies and
discussed its connection with the small x-behaviour of various types of
parton densities.
This model is based on a softening of the
mini-jet cross-sections due to  an s-dependent 
b-distribution in
the proton, which we calculate using a soft gluon resummation model down to
zero energy of the soft gluons.
\section*{Acknowledgments}
We acknowledge  partial support from EU CT-2002-0311. 

\end{document}